# Finite size effects in determination of thermal conductivities: comparing molecular dynamics results with simple models.

**Patrice Chantrenne[1,2], Jean-Louis Barrat[2]**

[1] Thermal Center of Lyon (CETHIL), UMR 5008 CNRS, INSA, Bât. S. Carnot, 20 Av. A. Einstein, 69621 Villeurbanne Cedex, France

[2] Laboratoire de Physique de la Matière Condensée et Nanostructures, UMR 5586 CNRS, UCBL, Bât. L. Brillouin, 43 Bd. du 11 Nov. 1918, 69622 Villeurbanne Cedex, France

**Abstract**

The thermal conductivity of nanometric objects or nanostructured materials can be determined using non equilibrium molecular dynamics (NEMD) simulations. The technique is simple in its principle, and resembles a numerical guarded hot plate experiment. The 'sample' is placed between a hot source and a cold source consisting of thermostatted sets of atoms. The thermal conductivity is obtained from the heat flux crossing the sample and the temperature profile in the system. Simulations results, however, exhibit a strong dependence of the thermal conductivity on the sample size. In this paper, we discuss the physical origin of this size dependence, by comparing MD results with those obtained from simple models of thermal conductivity based on harmonic theory of solids. A model is proposed to explain the variation of the thermal conductivity with system size.

**Nomenclature**

- $a_0$     lattice parameter (m)
- C     volume specific heat (J.K$^{-1}$.m$^{-3}$)
- D     density of mode
- e     film thickness (m)

| | |
|---|---|
| h | Planck constant (J.s) ($\eta = h/2\pi$) |
| $k_b$ | Boltzmann constant (J.K$^{-1}$) |
| K | wave vector (m$^{-1}$) |
| l | section characteristic lenght (m) |
| L | largest dimension of a cube, a wire and film (m) |
| m | masse (kg) |
| p | polarization |
| t | time (s) |
| T | temperature (K) |
| v | sound velocity (m.s$^{-1}$) |
| V | volume (m$^3$) |
| $\Delta t$ | time step (s) |
| $\varepsilon$ | minimum energy of the Lennard Jones potential (J) |
| $\lambda$ | thermal conductivity (W.m$^{-1}$.K$^{-1}$) |
| $\Lambda$ | wave vector largest path in a system (m) |
| $\omega$ | angular frequency (s$^{-1}$) |
| $\sigma$ | zero energy distance of the Lennard Jones potential (m) |
| $\tau$ | relaxation time (s) |

## I. Introduction

As with other transport or thermodynamic properties, the thermal behaviour of nanostructured materials or nanoelectronic devices cannot be simply inferred by extrapolating macroscopic behaviour to small scales. Instead, when the typical size of the device becomes comparable to interatomic distances, a discussion of properties and models at the atomic scale becomes essential. The total conductance of nanowire or nanowire arrays, superlattices, thin films and periodic thin film structures will depend on the thermal conductivity of each component and on the thermal resistance between them, all of which will be scale dependent. The aim of experimental and theoretical studies is to predict or to measure these characteristics [1-14]. However, the best experimental resolution is still larger than 100 nm. At smaller scales, atomistic numerical simulation appears to be an appropriate tool to predict the thermophysical properties.

Using atomistic modeling, three methods are available for computing thermal conductivities. The first one, mostly used for bulk systems, is equilibrium molecular dynamics (MD) [15, 16] using the appropriate Einstein or Green–Kubo relations. This approach can be extended to the determination of contact resistances [17]. An alternative route is to use non-equilibrium molecular dynamics (NEMD) [18, 19], and is more appropriate for inhomogeneous systems. The last approach combines a microscopic determination of phonon dispersion relations with a transport theory of the Landauer-Buttiker-type (which is generally used to calculate electronic transport properties) [20].

In this paper, we present NEMD simulations of heat transfer in finite size structures. The technique mimics numerically a guarded hot plate experiment. The 'sample' is placed between a hot source and a cold source consisting of thermostatted sets of atoms. The thermal conductivity is obtained from the heat flux crossing the sample and the temperature profile in the system. NEMD is therefore well adapted to study the influence of structural defects and solid interfaces. We show that the simulation results exhibit a strong dependence on sample size, and also on the type of boundary conditions. In order to understand the physical origin of these dependences, we compare our simulation results with a simple approach based on phonon transport theory.

The type of effects we are interested in are generic in nature, and our study a methodological one. Therefore no attempt was made to model a specific material in a realistic manner. Rather, we favour computational efficiency by using a simplistic model of particles interacting through a classical, pairwise, Lennard-Jones potential:

$$E_p(r) = 4\varepsilon \left[ \left(\frac{\sigma}{r}\right)^{12} - \left(\frac{\sigma}{r}\right)^{6} \right]$$

All the results in the following will be given in Lennard-Jones units. $\varepsilon = 1$, $\sigma = 1$ and $m = 1$. For the calculations, a modified version of the parallel MD code LAMMPS [21] was used.

Finally, we want to emphasize that our calculations are purely classical, which limits their applicability to cases where heat transport is essentially phononic, i.e. insulators or bad conductors.

The paper is organized as follows: in the next section, we describe our NEMD results, and how the size of the system and that of the thermostatted zones influence the thermal conductivity. Section III describes a semi-analytical approach used to rationalize our results, and section IV summarizes our conclusions.

**II. Size effects on simulation results**

In this section, we discuss two possible origins of the size dependence of thermal conductivity observed in our simulations, namely the influence of the thermostats (section II.1) and of the boundary conditions (section II.2). All the simulations are performed in a cubic simulation cell, containing a FCC lattice of Lennard-Jones atoms at mass density $\rho = 1.075 \, m/\sigma^3$. The side L of the cube was varied between $10a_0$ and $80a_0$ with $a_0 = 1.5496 \, \sigma$.

*II.1. Influence of the thermostatted zones*

Several papers describe the methodology used to simulate heat transfer [13, 22-26] with NEMD. A hot thermal reservoir and a cold thermal reservoir are used to create a temperature gradient in the system. To create these hot and cold region it is possible to control either the temperature or the energy given or taken from the thermal reservoir. Two important characteristics of the reservoirs must be well understood in order to avoid an incorrect interpretation of the simulation results:

- The heat source and the heat sink cannot be considered as classical boundary conditions of a continuous medium. They are part of the system, so that phonons modes are characterized by the whole dimension of the system and not by the size of the intermediate zones.

- As a heat flux requires a temperature gradient, if the thermal reservoirs were isotherm then no heat flux could flow to or from them. So, there is a temperature gradient within the thermal reservoirs, which depends on the heat flux.

Our thermal reservoirs are thermostatted slabs of thickness D, located at positions $-D/2 < x < D/2$ and $L/2 - D/2 < x < L/2 + D/2$. D varies from $2a_0$ to $30a_0$ depending of the dimension L of the cube. As mentioned above, a temperature gradient must exist within these zones. The thermostat simply maintains the average temperature of the reservoir constant [15, 16]. In order to control the temperature and to calculate the flux given to the heat source and taken from the heat sink efficiently, we use simple velocity rescaling: at the end of each time step, the temperatures of the heat source and heat sink are calculated and their velocity fields are rescaled in order to keep their kinetic energy constant and equal to the one corresponding to the target temperature [14, 25].

It is usually believed that this method modifies the thermal equilibrium at each time step. To limit this phenomenon and to allow natural temperature fluctuations, the rescaling can be done periodically and only when the instantaneous temperature is much departed from the target temperature. In fact, the kinetic energy correction is minimized by a systematic rescaling (at each time step and whatever the temperature value) which also allows local temperature fluctuations since the set of atoms is large enough. However, the velocity rescaling method may have a significant effect on the phonon population and on the local thermal equilibrium, which would modify the heat transfer in the system. To check for possible influence of the thermostats on the microscopic dynamics, the following quantities were investigated.

-the effective density of modes (i.e. the Fourier transform of the velocity autocorrelation of a particle) has been determined for a system at equilibrium at temperature $T=0.215\varepsilon/k_b$. It can be compared to the density of modes of a system with a heat source and heat sink such that the average temperature of the system is also equal to T. The two curves are superimposed in figure 1, where one can check that the influence of thermostats on vibrational dynamics is negligible.

-the local velocity distribution function has been determined in a system during a heat transfer simulation. The temperature profile in the system during this simulation is shown in figure 2. As expected, the temperature in the thermostatted zones is not homogeneous. For each atomic layer, the velocity distribution function is compared to the equilibrium Maxwell distribution function at the same temperature. Two such distribution functions are presented in figure 3, for layers taken in the middle of the heat source and in the middle of the intermediate (not thermostatted) zone. The two functions are different due to the temperature difference, but the numerical distribution exactly matches a Maxwell-Boltzmann distribution.

Finally, we also checked that the thickness of the thermostatted slabs does not influence our results (see figure 4) by comparing results for the thermal conductivity obtained with two different thickness (D = $10a_0$ and D = $15a_0$ for L = $50a_0$). However, the results exhibit a large dispersion if the number of free atoms between the thermal reservoirs is too large because the system then requires more time to reach the thermal local equilibrium.

Our conclusion is therefore that the thermostats do not significantly modify the microscopic dynamics of our system, and are therefore appropriate for performing heat transport simulations.

*II.2. Influence of the boundary conditions*

In MD simulations, the system is placed in a simulation box with a finite size. To predict bulk properties, the standard approach is to use periodic boundary conditions (PBC), in which the simulation box is periodically repeated in each direction [15, 16]. The system is finite, but completely homogeneous.

To study properties of nanometric structures free surfaces are used in all directions for a 3D object, in 2 directions for a wire and in one direction for a film. The object is embedded in a much larger simulation cell. Studying solids under these conditions can be difficult since evaporation can result in unwanted shape changes. To prevent the formation of the vapor phase, the system can be delimited by either with a set of fixed ("dead") atoms (hard wall) or with "phantom" atoms, i.e. weak harmonic springs that bind the atoms located on the system surface to their equilibrium position. The spring constant is chosen such that the associated vibrational frequency is smaller than the Einstein frequency, so as to minimize the influence of this constraint on vibrational dynamics.

In both cases, two phenomena can explain the size dependence of the thermal conductivity:
- the phonon number decreases with the system size and the wave vector distribution function can no longer be considered continuous; this "mode counting" effect is in principle similar for periodic boundary conditions and systems with free boundaries.

- if the phonon mean free path is larger than the box size then phonons will be more frequently scattered. This is quite obvious for systems with free surfaces, but for systems with periodic boundary conditions, this effect has not been clearly pointed out [29, 33, 34].

For a temperature $T=0.215\varepsilon/k_b$, the thermal conductivity of a cube of dimension $L=Na_0$ has been determined using NEMD. The influence of the cube size has been studied for $N=10$ to 80. Four kinds of boundary conditions were used: periodic boundary surfaces in the three directions, free surfaces with fixed ("dead") atoms, free surfaces with phantom atoms in all three directions, and unconstrained free surfaces. Due to the evaporation problem mentioned above, only one reliable data point could be obtained for the latter case. Figure 4 compares the results for the thermal conductivity obtained with these different boundary conditions, as a function of system size. Free surfaces with dead or phantom atoms yield, within numerical accuracy, the same conductivity. The thermal conductivity is slightly smaller for the periodic boundary conditions than for free surfaces. This would indicate that, somewhat surprisingly, *periodic boundary conditions result in stronger phonon scattering than free surfaces.*

Plotting the inverse of the thermal conductivity as a function of the inverse of the system size results in a linear relationship, both for free and periodic boundary conditions. The bulk thermal conductivity is obtained from the extrapolation of the regression line to system of infinite length (figure 5). The difference between the two asymptotic values is less than ten percent. Using the potential parameter of Argon($\varepsilon/k_b=120K$, $\sigma=0.34nm$) the bulk thermal conductivity is found to be equal to 1.4 +/-10% W/(mK) at 25 K, which is in good agreement with the value found in the literature (1.2 W/mK at 30 K increasing when the temperature decreases) [32]. The minimum cube size that would be required to obtain the bulk thermal conductivity with an accuracy greater than 3% is determined using the extrapolation curve: it is equal to $1170a_0$ for a system with periodic boundary conditions and $590a_0$ for a system with free surfaces. These systems are much too large (more than $10^9$ atoms) to be considered for MD, which justify the use of smaller systems and the extrapolation procedure.

## III. Simple models for the size dependence of the thermal conductivity

*III.1. Analytical expression of the thermal conductivity*

In a system of volume V at temperature T, the number of phonons with wave vector K and polarization p is given by the Planck distribution function [27, 28]:

$$\langle n_{K,p} \rangle = \frac{1}{(e^x - 1)} \quad (1)$$

with $x = \frac{\eta \omega(K,p)}{k_b T}$ and where $\omega(K,p)$ is given by the dispersion curve.

The internal energy of the phonons (K,p) per unit volume is the product of the number of phonons and their individual energy [27, 28]:

$$U(K,p) = \frac{\eta \omega(K,p)}{V(e^x - 1)} \quad (2)$$

which leads to its volume specific heat:

$$C(K,p) = \frac{\partial U(K,p)}{\partial T} = k_b x^2 \frac{e^x}{V(e^x - 1)^2} \quad (3)$$

The total volume specific heat is then equal to the sum over all the wave vectors and polarizations:

$$C = \sum_p \sum_k k_b x^2 \frac{e^x}{V(e^x - 1)^2} \quad (4)$$

If $\omega$ were constant then the expression for the volume specific heat would be the same as the one given by the Einstein model. Using elementary kinetic theory, the thermal conductivity associated with the phonons (K,p) can be written as [27, 28]:

$$\lambda(K,p) = C(K,p) v^2(K,p) \tau(K,p) \quad (5)$$

with: - v the group velocity of the phonons (K,p): $v(K,p) = d\omega/dK$,

- $\tau(K,p)$ is the phonon relaxation time.

Assuming a harmonic solid (no interaction between phonons), the phonons of different wave vectors and polarizations can be considered as a low density gas at constant volume. The total thermal conductivity is then the sum of the individual conductivities for all the wave vectors and polarizations:

$$\lambda = \sum_K \sum_p \lambda(K,p) = \sum_K \sum_p C(K,p)v^2(K,p)\tau(K,p) \qquad (6)$$

If the group velocity and the relaxation time are assumed to be constant, the classical expression for the thermal conductivity for an infinite system is obtained:

$$\lambda = v^2\tau \sum_K \sum_p C(K,p) = Cv^2\tau \qquad (7)$$

The thermal conductivity is constant and does not depend on the system dimension. So, to illustrate the influence of the system size on the thermal conductivity, it is necessary to take into account the actual variation of the group velocity and the relaxation time with the wave vector.

If the system size is infinite, the sum in (7) becomes an integral, which is most conveniently reexpressed using a change of variables from wavevector K to pulsation ω (through the dispersion curve and density of modes). The result for λ reads:

$$\lambda = \frac{\eta^2}{3k_bT^2V_{mol}} \int_0^{\omega_R} v(\omega)^2 \tau(\omega) D(\omega) \frac{\omega^2 e^{\eta\omega/k_bT}}{\left(e^{\eta\omega/k_bT}-1\right)^2} d\omega \qquad (8)$$

This expression has been proposed by Rosemblum et al. [29], who used it with the assumption of constant group velocity, to compute thermal conductivities of diamond crystals. It was named the 'Phonon Spectrum' (PS) method since it requires the knowledge of the density of mode, $D(\omega)$, which can be determined previously from equilibrium MD. For a perfect crystal, a general expression representing the phenomenological variation of the relaxation time is used:

$$\tau^{-1}(\omega) = \tau_U^{-1}(\omega) + \tau_{BC}^{-1}(\omega) \qquad (9)$$

with the relaxation time due to the Umklapp processes [28, 30]:

$$\tau_U^{-1}(\omega) = AK^2 T^\xi \exp(-B/T) \qquad (10)$$

and the relaxation time due to the presence of the system boundaries:

$$\tau_{BC}^{-1}(\omega) = v/(\alpha \Lambda) \qquad (11)$$

An other empirical formula [35] was proposed for $\tau_{BC}$:

$$\tau_{BC}^{-1}(\omega) = v*(1-s)/\Lambda \qquad (12)$$

where s represents the fraction of all phonon specularly scattered from the boundary surfaces ($s \in [0-1]$). Here Λ is the distance traveled between two scattering events by the boundaries.

Rosemblum et al. [29] assume $\Lambda$ to be constant and equal to the characteristic length of the system. A, B, $\xi$, $\alpha$ are identified in order to fit the calculated thermal conductivity to an experimental value. The PS method allows the prediction of the thermal conductivity of an infinite medium (L infinite leads to $\tau_{BC}^{-1}(\omega) = 0$), or of a finite object with characteristic length L in the three directions of space from the knowledge of the density of state and a few experimental values. As shown in the previous section, the density of modes can be calculated from simple molecular dynamics simulation at equilibrium. This method requires much less CPU time than NEMD and the results do not depend on the way heat transfer is simulated. However, the continuous integral over the angular frequency range implies that:

- the phonon properties do not depend on the wave vector direction. Since $\Lambda$ is assumed constant for all wave vectors, it is then not possible to study the size effects in nanowires or nanofilms. For such structures, $\Lambda$ is a function of the wave vector direction and can be considered infinite in one or two directions.

- the number of modes is infinite. The number of mode is equal to 3N-3 where N is the number of atoms in the system. If the system size is small, the sums in equation 6 can not be accurately transformed into a continuous integral.

To calculate the thermal conductivity of small systems, it seems therefore preferable to use equation 6 which contains more information on the wave vector: the number of modes, the group velocity, the relaxation time, the angular properties and their variation with the wave vector direction. This should in principle allows for proper inclusion of size and geometry effects, both in the mode counting and in the scattering. This method will be described in the following as the Wave Vectors (WV) method.

*III.2. Vibrational properties of Argon*

The thermal conductivity has been calculated with the WV method (equation 6) for a cube made of solid argon in order to compare the WV model results to the NEMD results for periodic boundary conditions and fixed boundary surfaces. For the periodic boundary conditions, two other geometries are studies: a nanofilm and a nanowire. To implement equation 6, vibrational properties of Argon must be known.

Solid argon has an fcc structure which unit cell is described by the three primitive vectors: $a_1(0, a_0/2, a_0/2)$, $a_2(a_0/2, 0, a_0/2)$ and $a_3(a_0/2, a_0/2, 0)$ where $a_0$ is the lattice parameter of the conventional fcc cell. The first Brillouin zone of the fcc structure is a cube truncated on its eight corners [27].

For a fcc crystal of dimension ($N_x a_0$, $N_y a_0$, $N_z a_0$) in the x, y and z directions, with periodic boundary conditions in the three directions, the wave vectors describing the crystal vibrations are linear combination of:

$$K_x = 2p_x\pi/(N_x a_0), \ K_y = 2p_y\pi/(N_y a_0), \ K_z = 2p_z\pi/(N_z a_0)$$

with $p_x \in \left]-N_x, N_x\right]$, $p_y \in \left]-N_y, N_y\right]$, $p_z \in \left]-N_z, N_z\right]$ (the wave vector are in a cube)

and $-3\pi/a_0 < K_x + K_y + K_z \leq 3\pi/a_0$, $-3\pi/a_0 < K_x - K_y + K_z \leq 3\pi/a_0$,

$-3\pi/a_0 < K_x + K_y - K_z \leq 3\pi/a_0$, $-3\pi/a_0 < -K_x + K_y + K_z \leq 3\pi/a_0$ (These later conditions allow to truncate the cube on its eight corners).

The number of wave vectors is equal to $4N_x N_y N_z$ which is the number of atoms. To simulate a cube, $N_x = N_y = N_z$, for a nanofilm : $N_x$ is finite and $N_y$ and $N_z$ tend to and infinite value. For a nanowire : $N_x = N_y$ and $N_z$ tends to an infinite value.

For a cube with dimensions ($N1 a_0$, $N1 a_0$, $N1 a_0$) and fixed boundary surfaces, the wave vectors describing the crystal vibrations are linear combination of:

$$K_x = p_x\pi/(N1 a_0), \ K_y = p_y\pi/(N1 a_0), \ K_z = p_z\pi/(N1 a_0)$$

with $p_x, p_y, p_z \in \left[1, 2(N1-2)\right]$ and $K_x + K_y + K_z \leq 3\pi/a_0 - 2/(N1-1)$

Due to the fixed boundary surfaces, the wave vectors are confined in 1/8 of the Brillouin zone of the fcc structure. However, the number of wave vectors is still equal the number of atoms : $4N1^3$ in this case. To compare the thermal conductivity of the cube with fixed boundary surfaces to the thermal conductivity of a cube with periodic boundary conditions, one has to set $N_x = N_y = N_z = N1-2$.

The dispersion curves of Argon should be known in order to calculate the angular and the group velocities for each wave vector. Assuming an harmonic crystal, there are one longitudinal mode and two degenerate transverse modes for each wave vector direction, which can be model as:

$$\omega(K,t) = \omega_{Mt} \sin(\frac{K\pi}{K_M 2}) \quad V(K,t) = V_{Mt} \cos(\frac{K\pi}{K_M 2}) \text{ for the two transverse modes and}$$

$$\omega(K,l) = \omega_{Ml} \sin(\frac{K\pi}{K_M 2}) \quad V(K,l) = V_{Ml} \cos(\frac{K\pi}{K_M 2}) \text{ for the longitudinal mode.}$$

For simplicity, it is also assumed that the dispersion curves are the same for all wave vector directions, with $K_M$ the maximum wave vector length in the first Brillouin zone. Considering the data for Argon in the literature [27, 36], the maximum angular and group velocities are equal to:

$$\omega_{Ml} = 1.35 \; 10^{13} \, rad/s \text{ and } v_{Ml} = 1800 \, m/s \text{ for the longitudinal mode and}$$

$$\omega_{Mt} = 0.6 * \omega_{Ml} \text{ and } v_{Mt} = 0.6 * v_{Ml} \text{ for the two transverse modes.}$$

Using equations 9, 10 and 12, the relaxation time at constant temperature can be written as :

$$\tau^{-1}(K) = A_1 K^2 + v*(1-s)/\Lambda(K)$$

For a cube, $\Lambda(K)$ is equal to the characteristic length of the cube. But, for the wire and the film, $\Lambda(K)$ is defined as the maximum length a wave can travel in the system between two boundaries. It is then calculated for each wave vector direction. The constant $A_1$ is determined so that the asymptotic value of the thermal conductivity calculated with the WV model for an infinite length is equal to the experimental thermal conductivity of Argon. At T= 25 K: $\lambda_{bulk} = 1.4 \, W/(mK)$ leads to $A_1 = 1.8921 \; 10^{-12} \, m^2/s$. All the results are normalized by this bulk value.

*III.3. Results*

The thermal conductivity is calculated for a cube of Argon at 25 K for the two limiting values of the scattering parameter s (figure 6):

- s = 1 : only specular reflexions occur at the boundaries. The phonon mean free path is not limited by the system boundaries. This condition is used to study the influence of the number of wave vectors on the thermal conductivity, independently of the system geometry. This is of course an artificial construction, which cannot be directly compared to MD simulations. The thermal conductivity of this cube made of an fcc solid Argon is smaller than $0.97 * \lambda_{bulk}$ for a dimension less than 50 $a_0$ for the periodic boundary conditions and 135 $a_0$ for the fixed boundary surfaces. Above these dimensions, mode counting effects can reasonably be neglected. The thermal conductivity of the system with free surfaces is smaller than the one of the periodic system: for periodic boundary conditions, the first Brillouin zone is uniformly sampled while, for fixed boundary surfaces, wave vectors are not allowed on the limit of the first Brillouin zone.

- s = 0 : phonons are scattered at the system boundaries. The phonon mean free path decreases (equations 9 and 11) due to the finite size of the system. The length $\Lambda(K)$ is then equal to the maximum distance a plane wave could travel in the wave vector direction. This leads to a smaller thermal conductivity than in the bulk value. For NEMD simulations with periodic boundary conditions, phonons reaching a boundary surface re-enter the system through another surface, which can be considered as a scattering event. For a system with free surfaces, phonons reaching a boundary surface are reflected and scattered back into the system. Analytical results with scattering and NEMD results can then be compared. Qualitatively, the agreement is quite good between the two methods: the relative variations are of the same order of magnitudes and the thermal conductivity of the cube with fixed boundary condition is greater than the one of the cube with periodic boundary conditions. Quantitatively the variation of the thermal conductivity due to the boundary conditions is much larger with NEMD than with the WV model, and the asymptotic value is reached for smaller system sizes with NEMD than with the WV model. These differences can easily be explained by the following assumptions, made in the WV model, and which may in fact not be valid : (1) vibration modes are plane waves, (2) the system is harmonic and (3) the phonon properties do not depend on the wave vector and polarization. It has been verified that the relative variations of the thermal conductivity are almost the same if one considers different dispersion curves for transverse and longitudinal polarization. So assumptions (1), (2) are certainly the ones which constitute the most serious approximations. Note also that we have used the same value of the scattering parameter s for the two situations. By adjusting the value of s to an intermediate value, it would of course be possible to increase, in the WV model, the difference between fixed and periodic boundary conditions, so that the WV results would be qualitatively closer to the NEMD results.

Another important difference between MD and WV model lies in the specific heat. In molecular dynamics simulations, the energy of each vibration mode is equal to $k_bT$. This leads to a constant specific heat, $k_b$, for all the vibration modes. Actually, as phonons follow the Planck distribution function, the specific heat depends on the wave vector and the temperature (equation 3). This effect can never be taken into account in classical MD simulations. Using the WV model and assuming a constant specific heat for all wave vectors, it is clear that the equipartition of energy in MD simulations leads to an overestimation of the thermal conductivity; we have checked, however, that the relative variations with system size are similar if the classical formula, rather than the correct Planck distribution, is used for the specific heat. This is reasonable, since finite size effects are dominated by long wavelength phonons, for which the classical formula is most accurate. Moreover, at high temperature, the phonon specific heat (equation 3) tends to $k_b$, so that the error in MD simulations due to this classical treatment of lattice vibrations will decrease.

As the WV model qualitatively reproduces the size dependence of the thermal conductivity of a nanoparticle, we have extended it to study the size dependence of the thermal conductivity of a nanofilm and a nanowire. Only periodic boundary conditions are considered since the difference with fixed boundary surfaces is small. The scattering parameter s is equal to zero. Films and wires actually have two characteristic lengths:
- the thickness e and the characteristic dimension L in the (y, z) plane for the nanofilm (figure 7a),
- the characteristic length l of the wire section (in the (x, y) plane) and the length L of the nanowire in the z direction (figure 8a).

The thermal conductivities $\lambda(e,L)$ and $\lambda(l,L)$ of the film and the wire are calculated within the WV model for increasing values of L (figures 7b and 8b). On this figures, the thermal conductivity appears to diverge for large values of L, at fixed e and L. The divergence, which is barely noticeable for the thick films or wires, becomes evident when the lateral dimensions decrease. Such a divergence is typical of low dimensional systems [37], and results from the particular role of the wavevectors that are parallel to the longer dimension (i.e. those with $k_x = 0$ in the film, with $k_x = k_y = 0$ in the wire). For these wavevectors, boundary scattering is absent, and the sum over ($k_y, k_z$) (film) or $k_z$ (wire) yields a divergent integral in the limit of large L. In practical cases, this divergence (which, in our model, is actually slightly overestimated [38] will be cutoff either by the device dimension or by a typical distance between defects in the larger dimension.

In order to define a thermal conductivity for films and wires, we have somewhat arbitrarily chosen to compute the value of $\lambda(e,L)$ and $\lambda(l,L)$ for a value of $L = 2000 a_0$, corresponding to a typical device size of 1 micrometer. The results, shown in figures 7c and 8c, show that, for the same characteristic length (l and e), the thermal conductivity of the film is larger than the thermal conductivity of the wire. This is expected, since boundary scattering is stronger in wires. An interesting feature is the nonmonotonic behaviour with the lateral characteristic size. As e or l decrease, the conductivity first decreases as the scattering by boundaries becomes more efficient, but eventually increases again at small thickness, when the low dimensional character and the associated divergence prevail. This nonmonotonic behaviour illustrates the difficulties encountered in extrapolating data obtained over a limited range of sizes in the range of nanometric scales.

**IV. Conclusion**

NEMD is one of the possible tools for determining the thermal conductivity of a material on an atomic scale. Even if NEMD is widely used, its reliability has sometimes been questioned. In this paper, the influence of the thermal reservoir, the boundary conditions and the system size on the thermal conductivity is studied by use of NEMD simulations of a cube with an fcc structure and a Lennard-Jones potential.

It has been shown that, when the thermal reservoirs are thermostatted with a systematic velocity rescaling, the local velocity distribution function is the Maxwell distribution function and the density of mode remains the same as the one obtained for an equilibrium system. Moreover, the temperature gradient between the thermostats is linear. Thus, it is believed that the local thermal equilibrium is achieved everywhere in the system. The thermal conductivity can then be calculated from the temperature gradient and heat flux.

An analytical model of the thermal conductivity is developed in order to give qualitative explanations of the thermal conductivity variations with the system size. In this model, the total thermal conductivity is considered as the sum of the individual thermal conductivity of all the phonon modes. This is the Wave Vector method. The phonon properties required are determined from the literature data for Argon, in order to compare the WV model results to NEMD results. All the wave vectors should be characterized but, for simplicity, the thermal conductivity has been calculated assuming that phonon properties are the same for all the wave vector directions.

The system size dependence of the thermal conductivity is confirmed with our NEMD results. The bulk thermal conductivity can be recovered from the extrapolation of the simulation results for finite systems towards an infinite system size. Qualitatively, the analytical model shows that the size dependence of the thermal conductivity is due to the:

- discretisation of the wave vectors for small systems,
- influence of the phonon scattering the boundary surfaces.

Due to the surface scattering, the size dependence of the thermal conductivity is expected for the system with free surfaces. Moreover, it is shown that the thermal conductivity does not depend on the way free surfaces are modeled: hard wall, phantom atoms and really free surfaces. For the system with periodic boundary conditions, the thermal conductivity also exhibits a size dependence and values smaller than the ones obtained for the system with free surfaces.

The wave vector model was used to study the size dependence of the thermal conductivity of a nanofilm and a nanowire. As the film thickness or the section typical length of the wire decrease, the thermal conductivity decreases as expected. However, for small enough film thickness or wire section (typically less than 20 lattice parameters for solid argon) the thermal conductivity increases and can become larger than the bulk thermal conductivity. This is due to the vibrational behavior of the film and wire which then resembles the one of a 2D and 1D system, respectively, for which the thermal conductivity diverges with the system length.

**Acknowledgement:** this work was supported by the CINES (CNRS) and the CDCSP (Univ. Lyon 1) for computer facilities.

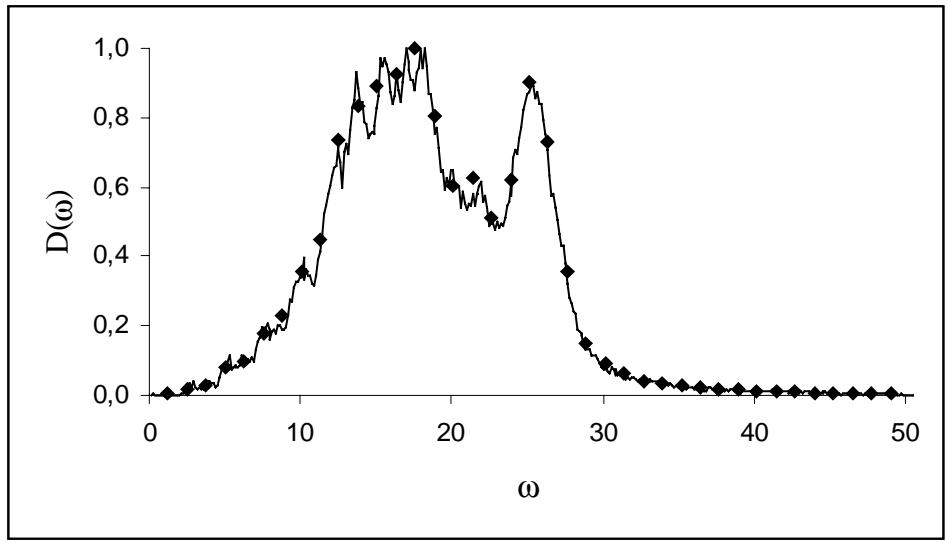

Figure 1: Density of mode D (in arbitrary units) versus angular frequency (Lennard-Jones units). Line : results for system at equilibrium; Diamonds : results for system with thermostatted zones.

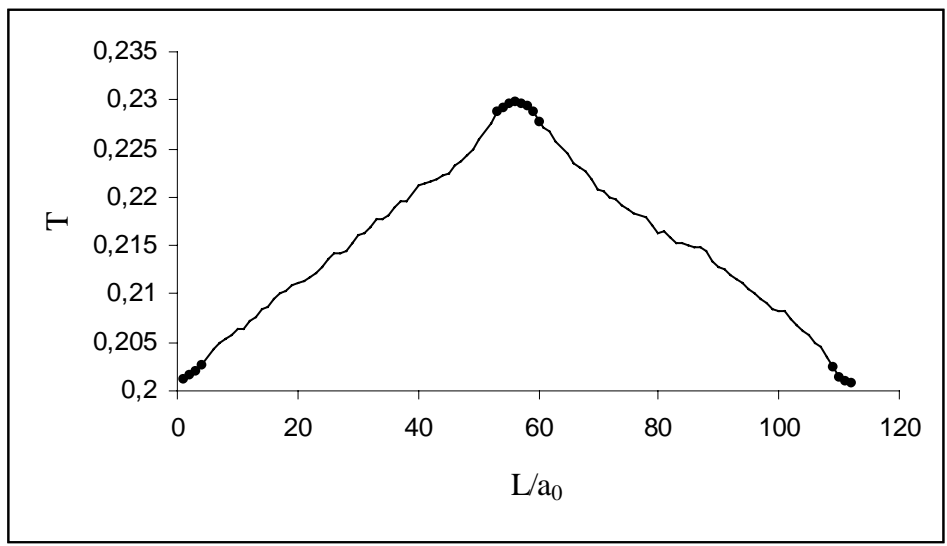

Figure 2: Temperature profile in a system with thermostatted zones as a function of the non-dimensional position in the direction of heat transfer. Periodic boundary conditions are used in all directions. The dots indicate the thermostatted zones.

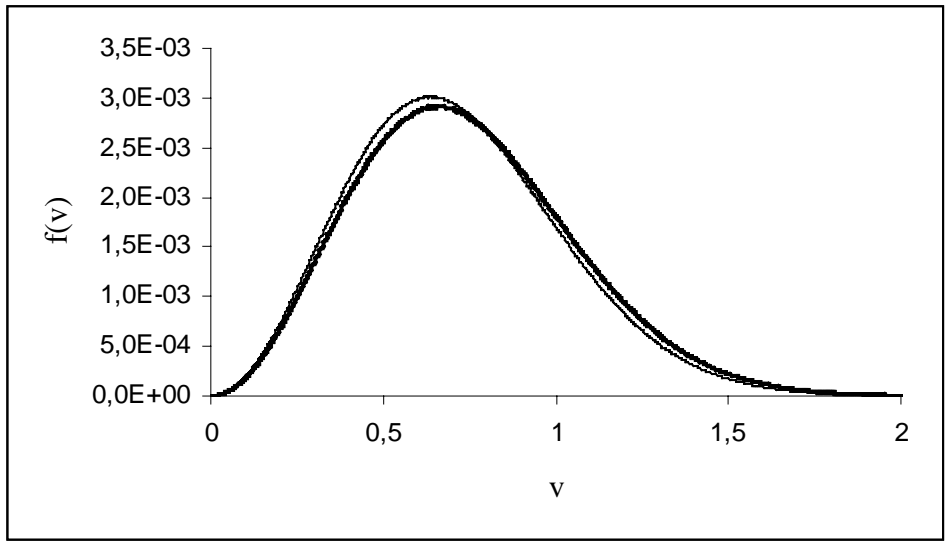

Figure 3: Velocity repartition function. Bold line: repartition function in the middle of the hot source. Thin line: repartition function in the middle of the intermediate bloc. In both cases, numerical and theoretical repartition functions are superimposed.

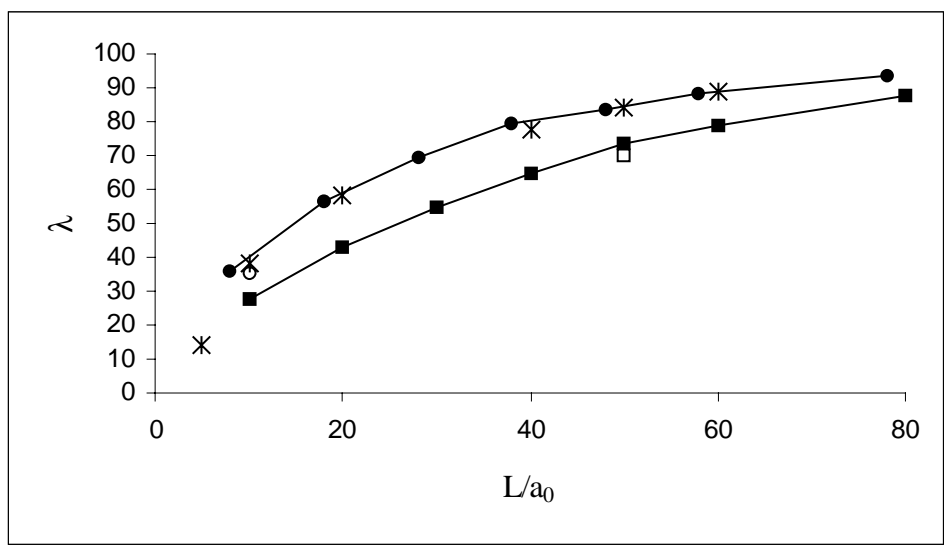

Figure 4: Thermal conductivity in LJ units as a function of the system size. Line with squares: periodic boundary conditions. Empty square: result for another thermostat dimension. Line with circles: free surfaces with dead atoms. Stars: free surfaces with phantom atoms. Empty circle: result for a system with really free surfaces.

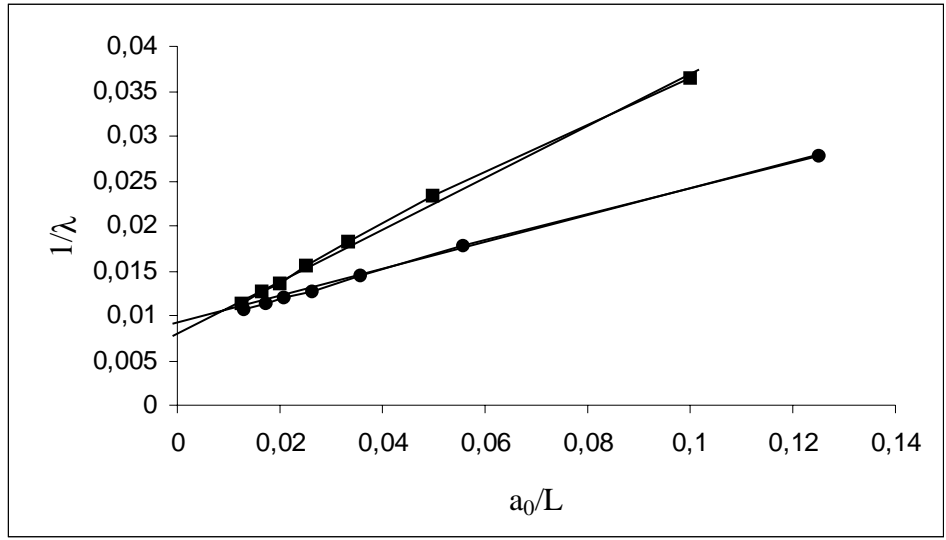

Figure 5: Inverse of thermal conductivity as a function of the inverse of the system size. The extrapolation of the regression line to an infinite system size gives the bulk thermal conductivity.

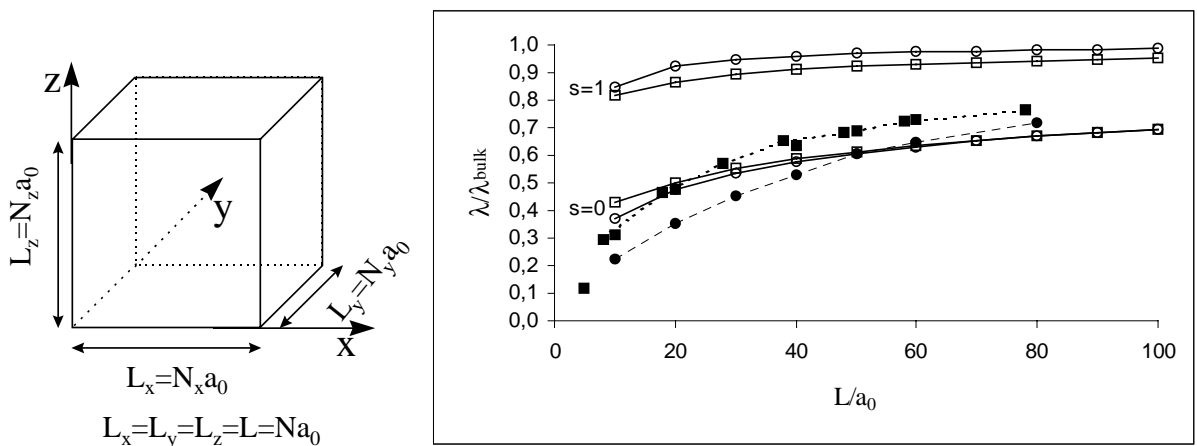

Figure 6: Normalized thermal conductivity of a cube as a function of its characteristic length. Comparison between the WV model (full lines) with NEMD (dashed lines) for periodic boundary conditions (lines with squares) and fixed boundary surfaces (lines with circles).

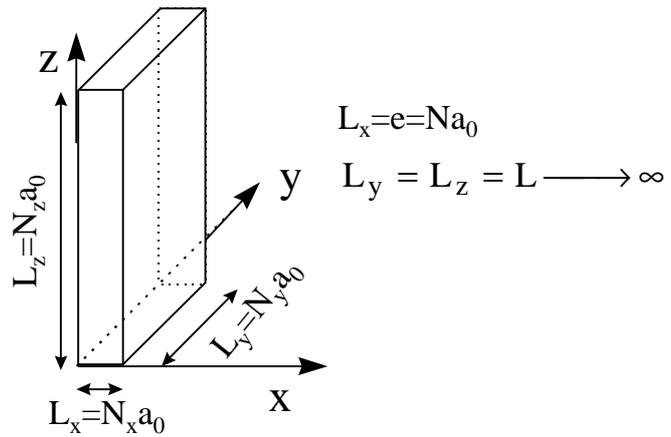

a

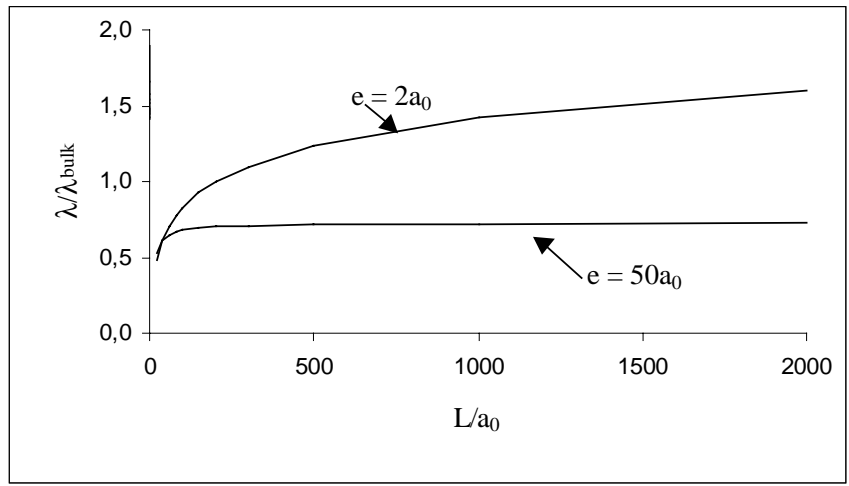

b

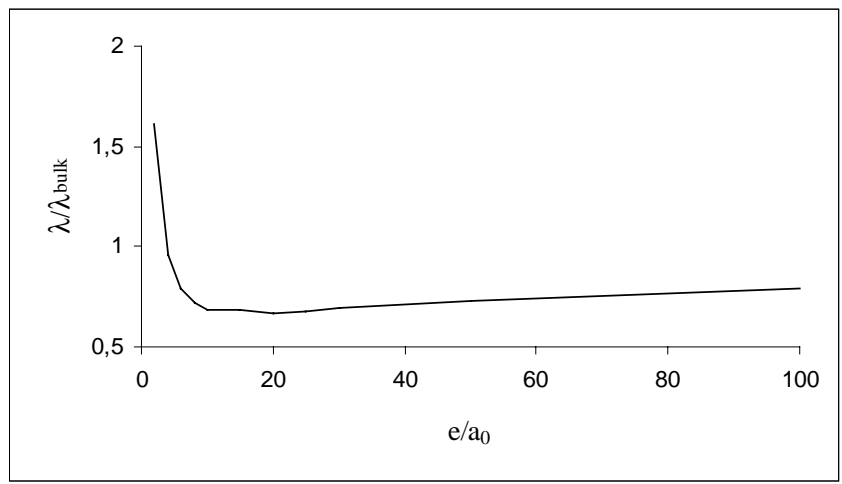

c

Figure 7: Normalized thermal conductivity of a film (a) as a function of L for two thickness and (b) as a function of its thickness for L=2000 $a_0$.

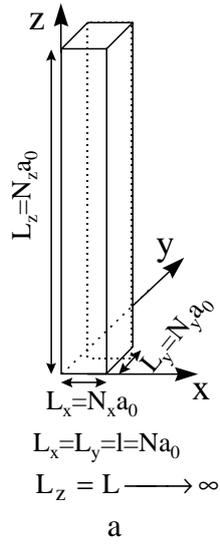

a

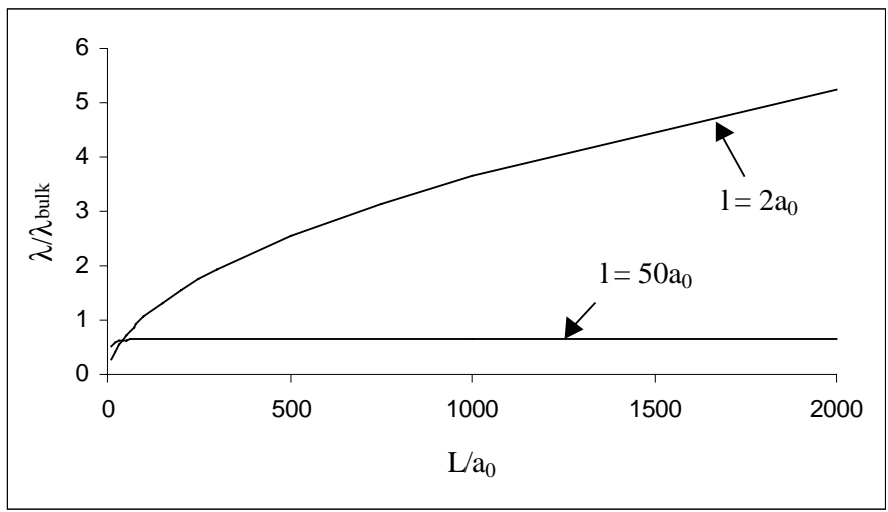

b

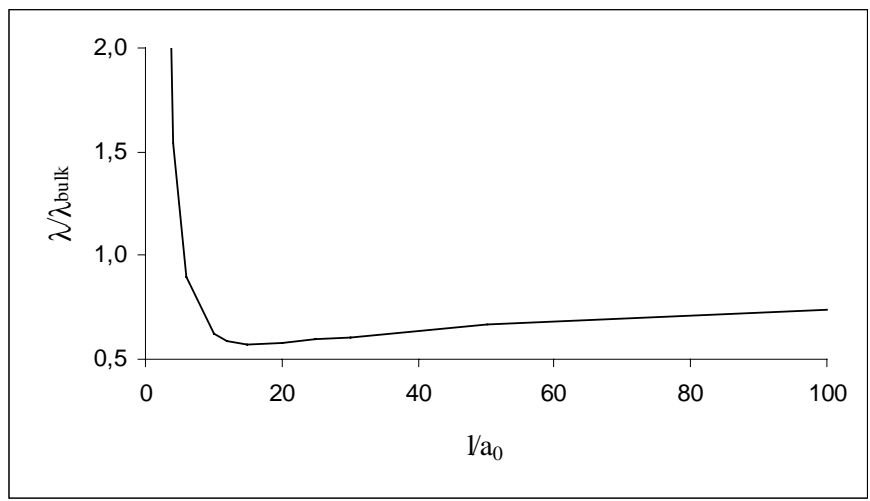

c

Figure 8: Normalized thermal conductivity of a wire (a) as a function of L for two values of l and (b) as a function of l for L infinite.